\documentclass[aps,prl,amsmath,amssymb,twocolumn,superscriptaddress]{revtex4}
\pdfoutput=1 
\usepackage[usenames]{xcolor}
\usepackage[pdftex]{graphicx}
\usepackage{amsmath, amstext, amssymb, amsfonts, amsxtra}
\usepackage{xspace}
\usepackage{blindtext}
\usepackage{here}
\usepackage[colorlinks=true,linkcolor=black,citecolor=blue,urlcolor=black]{hyperref}
\usepackage{siunitx}

\begin{document}
\title{Energy localization in interacting atomic chains with topological solitons}
\author{L. Timm}
\email[]{lars.timm@itp.uni-hannover.de}
\affiliation{Institut f\"ur Theoretische Physik, Leibniz Universit\"at Hannover, Appelstr. 2, 30167 Hannover, Germany}
\author{H. Weimer}
\affiliation{Institut f\"ur Theoretische Physik, Leibniz Universit\"at Hannover, Appelstr. 2, 30167 Hannover, Germany}
\author{L. Santos}
\affiliation{Institut f\"ur Theoretische Physik, Leibniz Universit\"at Hannover, Appelstr. 2, 30167 Hannover, Germany}
\author{T. E. Mehlst\"aubler}
\affiliation{Physikalisch-Technische Bundesanstalt, Bundesallee 100, 38116 Braunschweig, Germany}

\date{\today}

\begin{abstract}
Topological defects in low-dimensional non-linear systems
feature a sliding-to-pinning transition of relevance for a variety of research fields, ranging from biophysics to nano- and solid-state physics. 
We find that the dynamics after a local excitation results in a highly-non-trivial energy transport in the presence of a topological soliton, characterized by a strongly enhanced energy localization in the pinning regime.
Moreover, we show that the energy flux in ion crystals with a topological defect can be sensitively regulated by experimentally accessible environmental parameters. 
Whereas, third-order non-linear resonances can cause an enhanced long-time energy delocalization, robust energy localization persists for distinct parameter ranges even for long evolution times and large local excitations. 
\end{abstract}

\maketitle

Energy transport in low-dimensional systems attracts a sustained attention since the well-known results of Fermi, Pasta and Ulam in 1955~\cite{FPU}, which showed that ergodicity and thermalization are not inherent in anharmonic particle lattices. 
Subsequently, several low-dimensional lattice models have been employed to study microscopic energy transport~\cite{Savin2003,Lepri2003,Dhar2008}. The underlying dynamics of non-linear many-body systems can exemplarily be investigated by the Frenkel-Kontorova (FK) model~\cite{FK,BraunBook},an ubiquitous and paradigmatic model of nanofriction, which finds applications in a wide range of fields, from condensed matter physics~\cite{Brazda2018,Bohlein2011,Bruschi2002,Bak1982,Chaikin1995} to biophysics~\cite{Englander1980,Kuehner2007,Yakushevich2004}. It features a transition from a superlubric to a pinned phase, first described by Aubry~\cite{Aubry1983}.

Trapped ions constitute an excellent system to probe energy transport with atomic resolution in arbitrary dimensions both in the classical and in the quantum regime ~\cite{Ruiz2014,Bermudez2013,Toyoda2019}.
Recent theoretical works have shown that non-trivial transport features persist in one- and two-dimensional ion Coulomb crystals~\cite{Lin2011,Freitas2015}. 
First measurements have also monitored the time-resolved dynamics of an ion chain after a local excitation, revealing energy transport from one chain end to 
the other~\cite{Ramm2014,Abdelrahman2017}. 
Strong non-linearities become important when topological defects~(kinks) are induced into the system~\cite{Mielenz2013,Pyka2013,Ulm2013,Haljan2013,Kiethe2017}.
Kinks may be created by a quench due to the Kibble-Zurek mechanism~\cite{delCampo2010,Pyka2013} and the precise control of the kink via external trap parameters made it possible to investigate their properties~\cite{Landa2013,Partner2013}.

An example for the use of the good control of kinks in trapped ion systems is the study of nanofriction in the FK model.
Its original form may be emulated by superimposing an optical lattice to an ion chain~\cite{Bylinskii2016}, however, the same signatures of Aubry physics have been observed in a self-assembled and back-acting, two-dimensional crystal with defect~\cite{Kiethe2017}.
Experiments on both approaches showed the occurrence of reduced friction and a structural symmetry breaking at the Aubry transition~\cite{Aubry1983}.
\begin{figure}[t]
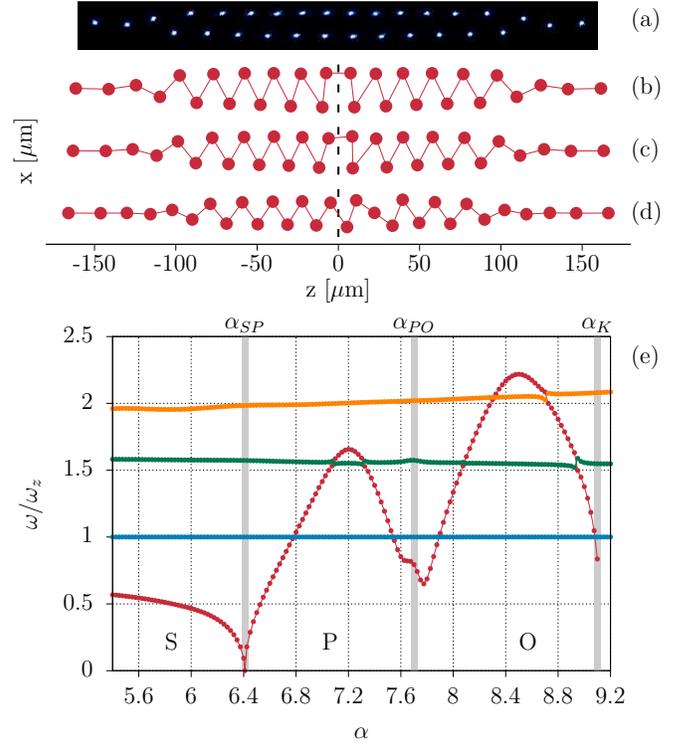

\begin{center}
\resizebox{\columnwidth}{!}{\large\input{tanjakink.tex}}
\resizebox{\columnwidth}{!}{\large \input{kink2.tex}}
\resizebox{\columnwidth}{!}{\large \input{kinkspectrum.tex}}
\vspace{-0.8cm}
\caption{(a) Ion chain with a central zigzag region with a kink in the sliding phase, experimental picture from Ref.~\cite{Kiethe2018}.
Computed equilibrium positions for sliding~(S) phase~(b), pinned~(P) phase~(c), and odd-kink~(O) phase~(d).
(e) $\alpha$-dependence of the frequency of the kink mode (red) and higher modes, normalized by $\omega_z$. The S-to-P and P-to-O transitions as well as $\alpha_\text{K}$ are marked by grey bars.}
\vspace{-1.0cm}
\label{kink}
\end{center}
\end{figure}
Other efforts have been made to investigate the transport properties of solitonic excitations~\cite{BraunBook,Oxtoby2007,Ahufinger2004,Abdullaev2010}. Whereas the breaking of a system symmetry, e.g. given by the external confinement in trapped ion systems, may cause a directed charge transport~\cite{Brox2017}, we focus on the effects of the autonomous symmetry breaking at the Aubry transition onto the dynamics of a local excitation.

\begin{figure*}[t]
\centering
\resizebox{2\columnwidth}{!}{\large\input{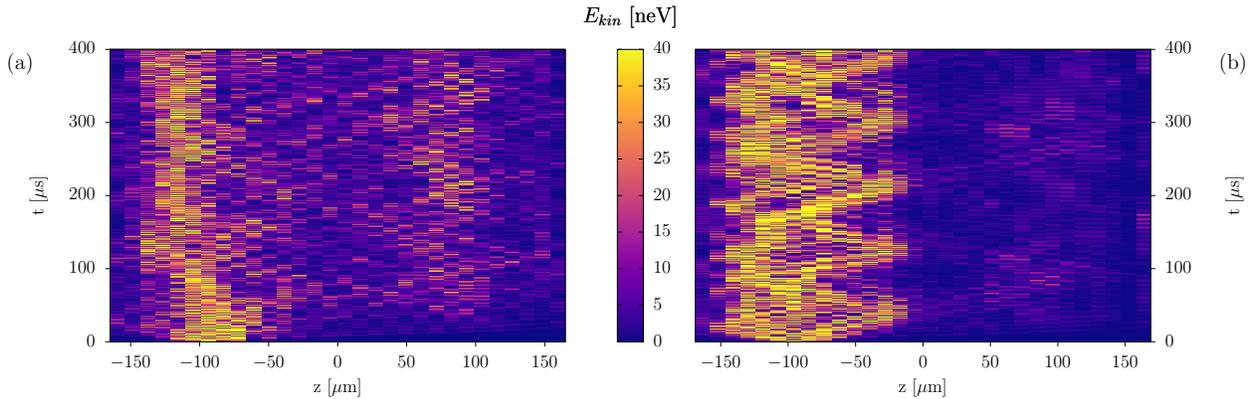}}
\vspace{0.3cm}
\caption{Dynamics of the kinetic energy after a {1}~\si{\micro\m} displacement of ion $7$ along $z$, calculated using Eq.~\eqref{harmApprox}, for 
(a) $\alpha= 5.5$~(sliding phase) and (b) $\alpha = 6.8$~(pinning phase).}
\vspace{-0.5cm}
\label{axialheatmap}
\end{figure*}
In this letter, we show that energy transport after a local ion displacement, as in Refs.~\cite{Ramm2014,Abdelrahman2017}, shows a tunable, highly non-trivial dynamics of energy 
in the presence of a topological soliton. 
Whereas the energy of a local excitation stays strongly localized in the pinned phase, the defect becomes transparent to energy flux in the sliding regime.

Moreover, third-order non-linear resonances between crystal modes result in a highly non-trivial dependence of the long-time dynamics on the aspect ratio of the transversal to the longitudinal trap confinement.
Whereas enhanced delocalization occurs for certain aspect ratios, slightly altered values lead to strong robust localization within the 
pinning regime for very long evolution times and large energy displacements. 
\paragraph{Model.--}
We consider $N$ ions placed on the $xz$ plane, with the crystal having a larger spread along the $z$ axis. 
The ions experience a confining potential that results from 
the sum of an external harmonic trap potential and the Coulomb interaction,
\begin{align}
V=\sum_i \frac{m\omega_z^2}{2}(z_i^2+\alpha^2 x_i^2)+\frac{1}{2}\sum_{i,j\neq i}
\frac{e^2}{4 \pi \epsilon_0} 
\frac{1} {|\vec{r}_i-\vec{r_j}| }
\label{Potential}
\end{align}
where $\vec{r}_i=(x_i,z_i)$ is the position of the $i$-th ion, $m$ the ion mass, $e$ the elementary charge, $\epsilon_0$ the vacuum permittivity, and 
$\omega_z$ the harmonic trap frequency along the $z$ direction. The parameter $\alpha=\omega_x/\omega_z$ characterizes the aspect ratio of the trap. 
In our numerical simulations below we consider a typical, experimental scenario, $N$ = 30 $^{172}\mathrm{Yb}^+$ ions and 
$\omega_z / 2\pi$ = {25}~\si{\kilo\Hz}.

\paragraph{Equilibrium.--} At equilibrium the ions are located at positions $\vec{r}_{i,0}$. For $\alpha>\alpha_{\text{ZZ,L}}$~($= 11.98$ in our case) the crystal remains a 1D chain along the $z$-axis. 
Below that value a zigzag phase (ZZ) develops at the chain center \cite{Fishman2008}. For $\alpha<\alpha_\text{K}$~($= 9.1$ in our case) Peierls-Nabarro barriers stabilize a kink 
inside the zigzag crystal by providing a local energy minimum \cite{Partner2013}. We constrain our analysis to $\alpha<\alpha_\text{K}$, since we 
are interested in the energy transport in the presence of a kink. To prepare a defect, we mirror one half of a zigzag crystal on the $z$-axis and let the kink slip into the local energy minimum given by the Peierls-Nabarro barriers.

At $\alpha_\text{SP}$~($= 6.4$ in our case) the kink undergoes an Aubry-type transition, between 
a sliding~(S) phase, characterized by mirror symmetry~(Fig.~\ref{kink}(b)) and a pinning~(P) phase, in which the symmetry is spontaneously broken~(Fig.~\ref{kink}(c)).
At $\alpha_\text{PO}$~($= 7.7$ in our case) the kink undergoes a crossover to the odd~(O) phase, in which the radial trap is sufficiently strong to force an ion in between the two chains~(Fig.~\ref{kink}(d))~\cite{Pyka2013}. As shown below, the onset of these kink transitions has dramatic consequences for the excitation transport in the crystal.

\paragraph{Harmonic approximation.--} Expanding the potential given in~\eqref{Potential} up to second order around the equilibrium positions $\{\vec r_{i,0} \}$, 
each ion experiences an effective harmonic oscillator potential with two degrees of freedom, $\mu = 1,2$. 
The excitations of these local harmonic oscillators (vibrons) are determined by the Hamiltonian
\begin{align}
H=\sum_{i,\mu}\hbar\omega_i^\mu{a_i^\mu}^\dagger a_i^\mu+\sum_{i,j\neq i}\sum_{\mu,\eta}t_{i,j}^{\mu,\eta}\left ({a_i^\mu}^\dagger a_j^\eta+\mathrm{h.c.} \right )
\label{harmApprox}
\end{align}
where ${a_i^\mu}^\dagger$~($a_i^\mu$) denotes the creation~(annihilation) of a $\mu$ vibron at ion $i$, $\omega_i^\mu$ is the frequency of a $\mu$ vibron at ion $i$, and $t_{i,j}^{\mu,\eta}$ characterizes the vibron hopping between ions $i$ and $j$. The oscillator 
and hopping energies depend mainly on $\{\vec r_{i,0} \}$, which in turn depend on $\alpha$ and $N$.

Another picture, leading to the same results, is considering global phonon-like excitations of the crystal.
The dynamics of an excitation in this description is determined by the dephasing of the amplitudes of these eigenmodes, which oscillate each with their respective frequency. 
 In particular, the presence of a kink results in a global mode with strongly localized amplitude at the defect~(kink mode), which shows an interesting frequency dependence on $\alpha$~(Fig.~\ref{kink}(e)), vanishing at the Aubry-type transition and showing a marked minimum at the onset of the odd phase.
 
\paragraph{Energy localization.--} We first determine the equilibrium positions of the ions for a given $\alpha<\alpha_K$~(we also restrict to $\alpha>5.5$, since for lower $\alpha$ 
the zigzag may undergo a structural transition into a three-layer arrangement). We then perform instantaneously at time $t=0$ a displacement 
of a single chosen ion, resulting in a coherent state, which may be experimentally realized as in Refs.~\cite{Ramm2014}. After this local excitation the system evolves freely, and the initially localized energy may propagate through the ion crystal. In order to study this propagation, we evaluate the kinetic energy $E_{i}(t) = m\dot{\vec{r}}_i(t)^2/2$ for the $i$ ion at time $t>0$. 

The excitation dynamics is markedly different in the S and in the P phases, as shown in Fig.~\ref{axialheatmap} for the case of an initial displacement of {1}~\si{\micro\m} of ion $7$ along the 
$z$ direction. Whereas in the S phase the energy can be quickly transported across the system within tens of \si{\micro\s}, in the P phase 
the kink strongly blocks energy transport. The excitation remains localized within one half of the crystal, unable to cross the defect region. 
The starkly different behavior can be traced back to the spatial amplitude distribution of the phonon-like modes that become populated by the initial excitation. 
Due to the emergence of new Peierls-Nabarro barriers at the Aubry-type transition, the kink is forced to a different local energy minimum, located away from the symmetry axis, and the crystals spatial symmetry is spontaneously broken~\cite{Partner2013}. 
Therefore, the amplitude of the dominant modes are strongly localized in the excited half of the crystal, which leads to an energy blockade.

Energy localization is best investigated by monitoring the time-averaged kinetic energy~(after a given time $\tau$) at one half of the crystal, e.g.
$E_\text{left}=\sum_{z_i < 0} \langle E_i\rangle$, with $\langle E_i\rangle=\frac{1}{\tau}\int_0^\tau \text{d}t E_{i}(t)$. We define
$\Delta E$ as the ratio between $E_\text{left}$ and the total energy of the system. Mirror symmetric energy distributions are characterized by 
$\Delta E= 0.5$, whereas energy localization at the left half results in $\Delta E> 0.5$.

\begin{figure}
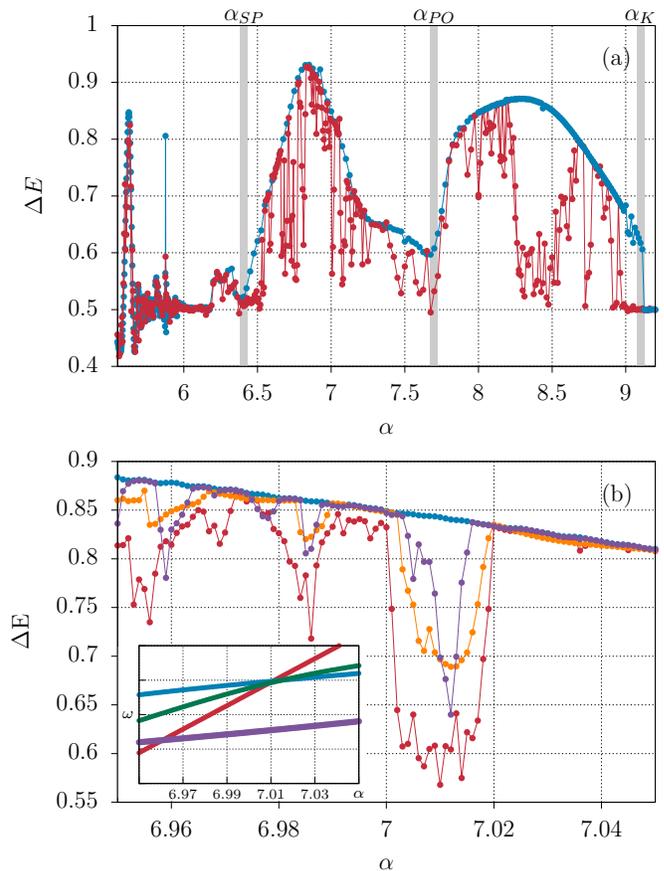

\centering
\resizebox{\columnwidth}{!}{\large\input{kinktransalpha.tex}}
\resizebox{\columnwidth}{!}{\large\input{kinktransalpharegion1.tex}}
\vspace{-0.7cm}
\caption{(a) $\Delta E$ as a function of $\alpha$ averaged for $\tau=${100}~\si{\milli\s} after an initial displacement of the leftmost ion by {1}~\si{\micro\m} along $x$. Blue circles indicate 
the results based on the harmonic approximation~(Eq.~\eqref{harmApprox}), whereas red circles depict the results obtained from a 
molecular dynamics simulation, considering the full potential~\eqref{Potential}. Grey bars indicate the same phase transitions 
of Fig.~\ref{kink}(e). (b) Same as (a), zooming in the region $6.95<\alpha<7.05$. Red and blue curves depict the same results as (a). The orange curve 
depicts the results for $\tau =$ {20}~\si{\milli\s}, whereas the purple curve shows $\Delta E$ for an initial displacement of {0.5}~\si{\micro\m}. 
We depict in the inset the mode frequency of the most excited mode by the initial displacement~(red), as well as the sum of two non-excited modes (blue, green and purple), showing 
the appearance of non-linear resonances.}
\label{transalpha}
\vspace{-0.5cm}
\end{figure}

\paragraph{Harmonic transport.-- } Figure~\ref{transalpha}(a) shows $\Delta E$, for an initial displacement of the leftmost ion by {1}~\si
{\micro\m} along $x$, averaged over $\tau=$ {100}~\si{\milli\s} (much longer than the dynamics timescale $\text{T}_z=2\pi/\omega_z=$ {40}~\si{\micro\s}). The blue circles depict our results using the harmonic approximation~(Eq.~\eqref{harmApprox}). 
For $\alpha<\alpha_{SP}$ the energy distribution is mainly symmetric~($\Delta E=0.5$) due to the transparency of the kink in the S phase. 
However, for some $\alpha$ values the energy remains localized even in the S phase, leading to sharp peaks in $\Delta E$.
The anomalously slow delocalization results from the crossing of two harmonic modes for certain $\alpha$ values. If these modes are 
strongly populated by the initial excitation, the mode degeneracy results in very slow dephasing, handicapping delocalization even at long times.

At $\alpha>\alpha_\text{SP}$, the crystal symmetry is spontaneously broken, leading to a rapid increase of $\Delta E$, which peaks at $\alpha\simeq 6.8$, for which 
over $90\%$ of the energy remains localized at the left side. Localization is again strongly enhanced at $\alpha_{PO}$ 
by the appearance of the odd kink, reaching up to a maximal $\Delta E=0.85$. Finally, at $\alpha_K$ the kink disappears, and with it the energy localization, resulting in $\Delta E = 0.5$.
Although the quantitative values of $\Delta E$ depend on the initially excited ion, the amplitude, and the direction of the displacement, similar significant localization 
is observed at the Aubry-type transition and at the onset of the odd kink for all choices of the initial excitation.

\paragraph{Anharmonic transport.--} In order to describe the actual energy transport, especially in the long time limit and for larger displacements, we must abandon the 
harmonic approximation. We hence employ molecular dynamics, considering the full potential~\eqref{Potential}. The corresponding results are depicted with red circles in 
Fig.~\ref{transalpha}(a). Anharmonic effects result in considerable delocalization that takes place on a timescale of hundreds of ms, slowly compensating the energy imbalance between the left and right halves. Interestingly, non-linear delocalization occurs at particular 
windows of $\alpha$ values, interspaced by regions that remain strongly localized and well described by the harmonic approximation for very long times, well over $0.1$ s. Figure~\ref{transalpha}(b), which zooms in the window $6.95<\alpha<7.05$, clearly 
illustrates the dense structure of harmonic~(localized) and anharmonic~(delocalized) regions.

The observed anharmonic delocalization of energy for particular $\alpha$ values is due to a third-order resonance of the phonon-like modes, which occurs when the frequency $\omega_{exc}$ of a eigenmode, localized in one crystal half and strongly populated by the initial excitation, equals the sum of two lower-lying mode frequencies which have a significant amplitude in both crystal halves~(see the inset in Fig.~\ref{transalpha}(b)). 
This leads to a strong coupling between these three eigenmodes, enabling transport across the kink.
Our results for different averaging times~(Fig.~\ref{transalpha}(b)) show that anharmonic delocalization occurs in a typical time scale 
of 10 - 100~\si{\milli\s}.

\begin{figure}
\centering
\resizebox{\columnwidth}{!}{\large\input{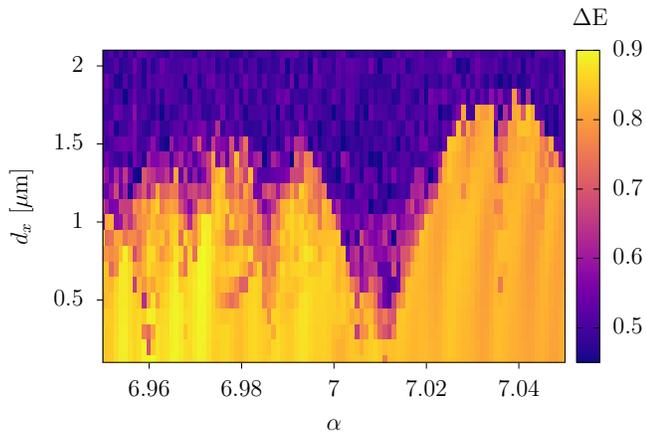}}
\vspace{-0.6cm}
\caption{$\Delta E$ for different initial displacement amplitudes $d_x$ of the leftmost ion along $x$, integrated over $\tau=100$ms.}
\label{transamplitude}
\vspace{-0.3cm}
\end{figure}

Larger initial displacements obviously enhance non-linear effects, leading to a broadening of the anharmonic resonances, as illustrated 
in Fig.~\ref{transamplitude}, where the observable $\Delta E$ is shown for different displacement amplitudes. 
Within the harmonic windows of $\alpha$ values, the localization is surprisingly robust up to very large displacements of {1.5}~\si{\micro\m}. For larger displacements 
the energy brought into the system is 
large enough to break through the energy blockade for all choices of $\alpha$. Figure~\ref{transamplitude} shows as well 
clear dips at the broadened non-linear resonances. In the vicinity of the resonance at $\alpha\simeq 7.01$ even tiny 
displacements result in non-linear delocalization due to mode coupling.

\paragraph{Conclusions.--} Ion Coulomb crystals with a topological kink soliton are characterized by a non-trivial excitation dynamics, which we have studied 
for the experimentally relevant case of an initial displacement of a single ion. Energy localization is very strongly enhanced at the Aubry-type transition, 
as a result of the presence of dominant localized phonon-like modes associated to the spontaneous symmetry breaking in the pinning phase. Strongly enhanced localization 
also results at the onset of the odd kink. Although the short-time evolution is to a large extent given by the harmonic approximation, the long-time dynamics presents 
a highly non-trivial dependence on the aspect ratio $\alpha$ of the external trap confinement. Whereas non-linear third-order resonances between phonon-like modes lead to reinforced delocalization for narrow windows of $\alpha$, non-resonant windows present robust energy localization even for very long evolution times and large initial displacements.
Overall, our results demonstrate the controllable, valve-like behavior of a defect, regulating the energy flux through a non-linear atomic crystal.
It should be noted that the qualitative features of the shown energy localization are only due to the existence of a solitonic defect with a sliding-to-pinning transition, placed into a quasi one-dimensional chain, which can be realized in various systems (e.g. by a loop defect in a DNA strain~\cite{Yakushevich2004}). 
Consequently, a consideration of other types of repulsive interactions should retrieve results similar to our case of an ion Coulomb crystal.

This work hence lays the basis for future studies of the transport of different kinds of quantum excitations and their interactions in non-linear many-body systems, which, in our case, can be switched on and off by slight changes of external trap parameters. 
Furthermore, our results are of interest for the cooling and the motional control of ion crystals in experiments dedicated to precision spectroscopy~\cite{Herschbach2012,Arnold2015,Aharon2019} and tests of fundamental physics~\cite{Sanner2019,Mehlstaeubler2018}.
They also emphasize the importance of defects for the exchange of energy between two thermal baths in low-dimensional systems, from solid-state nanosystems to biomolecules like DNA~\cite{Savin2003,Kuehner2007}, and path the way to future experiments on quantum thermodynamics~\cite{Lin2011,Bermudez2013,Freitas2015}. 

\acknowledgments
We thank J. Kiethe for discussions and comments on the manuscript.
The authors acknowledge the support from the DFG (SFB 1227 DQ-mat and EXC 2123 QuantumFrontiers) and the Volkswagen Foundation. This project has received funding from the European Metrology Programme for Innovation and Research (EMPIR) co-financed by the Participating States and from the European Union’s Horizon 2020 research and innovation programme (Project No. 17FUN07 CC4C).
\vspace{-12pt}
\bibliography{energytransport.bib}
\end{document}